%% file: fpcp04_zupan.tex
\documentclass[12pt]{article}
\usepackage{epsfig,graphicx}
\usepackage{fpcp04}
\usepackage{amsmath}

\input fpcp04-macro
\begin{document}

\Title{Determining $\alpha$ and $\gamma$ - theory}
\bigskip

%
\label{ZupanStart}
\renewcommand{\Im}{\operatorname{Im}}

%
\author{ Jure Zupan${}^{a,b}$\index{Zupan, J.} }

%
\address{
${}^a$ Josef~Stefan Institute, 
 39, P.O. Box 3000, 
1001 Ljubljana, 
Slovenia\\
${}^b$ Department of Physics, Carnegie Mellon University, Pittsburgh, PA 15213, USA
}

\makeauthor\abstracts{
In this short review presented at FPCP04, Daegu, Korea, we discuss methods leading to determinations of 
$\alpha$ and $\gamma$
with practically no theoretical error. The remaining theoretical errors due to isospin breaking, 
neglecting of electroweak penguins or coming from
other sources are addressed.
}

\section{Introduction}
We are entering a period of time, when direct determinations of the angles $\alpha$ and $\gamma$ of the standard unitarity 
triangle are becoming possible. In this talk we will review the methods that are used at present and the related theoretical
uncertainties. Surprisingly enough, some of the most useful methods were not even talked about before 2003. 
The questions that will be addressed are therefore
(i) what is the ultimate precision of different methods
and (ii) what are we learning about $\alpha, \gamma$   now?
The last question has been covered in great detail in talks by experimental colleagues \cite{Exp_talks}, so only the 
final results will be given here.

How can one measure $\alpha$ and $\gamma$? The sensitivity to the phases comes from 
interference. Useful methods thus rely on channels with at least two interfering amplitudes and/or interference
between mixing and decay. In order to extract the weak phases, however,  one needs to evaluate unknown hadronic parameters
that also enter the obsevables. A conservative approach to this problem is to extract all the hadronic parameters
from experiment. This is accomplished by   
using symmetries of QCD (e.g. $C$, $P$, isospin), and by
finding channels, where all parameters are obtainable from experiment. 
Another approach is to calculate the hadronic parameters using theoretical frameworks like QCD factorization, PQCD, and
SCET. This later avenue will not be exploited here and the reader is referred to  
\cite{Th_talks} for further details.

\section{Measuring $\alpha$}
\subsection{$B(t)\to \pi\pi$}
This method is due to Gronau and London and dates back almost 15 years ago \cite{Gronau:1990ka}. Let us review the method
step by step to see where the approximations enter. A completely general isospin decomposition of the decay amplitudes is
\begin{equation}
\begin{split}
A_{+-}&=\langle \pi^+\pi^-| H| B^0\rangle = -A_{1/2}+\tfrac{1}{\sqrt{2}} A_{3/2}-\tfrac{1}{\sqrt{2}}A_{5/2},\\
A_{00}&=\langle \pi^0\pi^0| H| B^0\rangle ~= \tfrac{1}{\sqrt{2}}A_{1/2}+ A_{3/2}\qquad - A_{5/2},\\
A_{+0}&=\langle \pi^+\pi^0| H| B^+\rangle =\qquad \qquad\tfrac{3}{2} A_{3/2}\quad + A_{5/2},
\end{split}
\end{equation}
where the notation for the reduced matrix elements is $A_{\Delta I}$. Equivalent relations hold for $\bar B^0$, $B^-$
decay amplitudes $\bar A_{+-}$, $\bar A_{00}$, $\bar A_{+0}$.
Note that the $\Delta I=5/2$ operators are not present in the effective weak Lagrangian, so that
$A_{5/2}$ can only arise from isospin breaking final state rescattering effects, such as $\Delta I=2$ 
electromagnetic rescattering of two pions. One can thus estimate 
 $A_{5/2} \sim \alpha A_{1/2}$. Setting $A_{5/2}=0$ therefore means neglecting a  $\sim 1 \%$ correction. 
Making this approximation one obtains two
 triangle relations
\begin{equation}
A_{+-}+\sqrt{2}\; A_{00}=\sqrt{2} \; A_{+0}, \qquad\qquad
\bar A_{+-}+\sqrt{2} \;\bar A_{00}=\sqrt{2} \; \bar A_{+0}.
\label{Zupantriangle-rel}
\end{equation}
Aside from possible electroweak penguin operator (EWP) contributions, $A_{+0}$ is a pure tree. 
Neglecting EWP one has an additional relation
\begin{equation}\label{Zupantree-rel}
e^{i\gamma} A_{+0}=e^{-i\gamma} \bar A_{+0} \quad~\Rightarrow ~\quad|A_{+0}|= |\bar A_{+0}|.
\end{equation}
This allows to extract $\sin 2\alpha$ from
$\Gamma(B^0(t) \to \pi^+\pi^-)\propto [ 1 + C_{\pi\pi}\cos\Delta mt - S_{\pi\pi}\sin\Delta mt]$
using the construction of Gronau and London \cite{Gronau:1990ka}. 
The observable $\sin( 2\alpha_{\rm eff})=S_{\pi\pi}/\sqrt{1-C_{\pi\pi}^2}$
${}$ is directly related to $\alpha$ through
$2\alpha=2 \alpha_{\rm eff}-2 \theta$,
where $\theta$ is defined on Fig. 1.

\begin{figure}[t]
\label{ZupanGLtriangle-fig}
\label{Zupanfig:pipi}
\begin{center}
\includegraphics[width=7.cm]{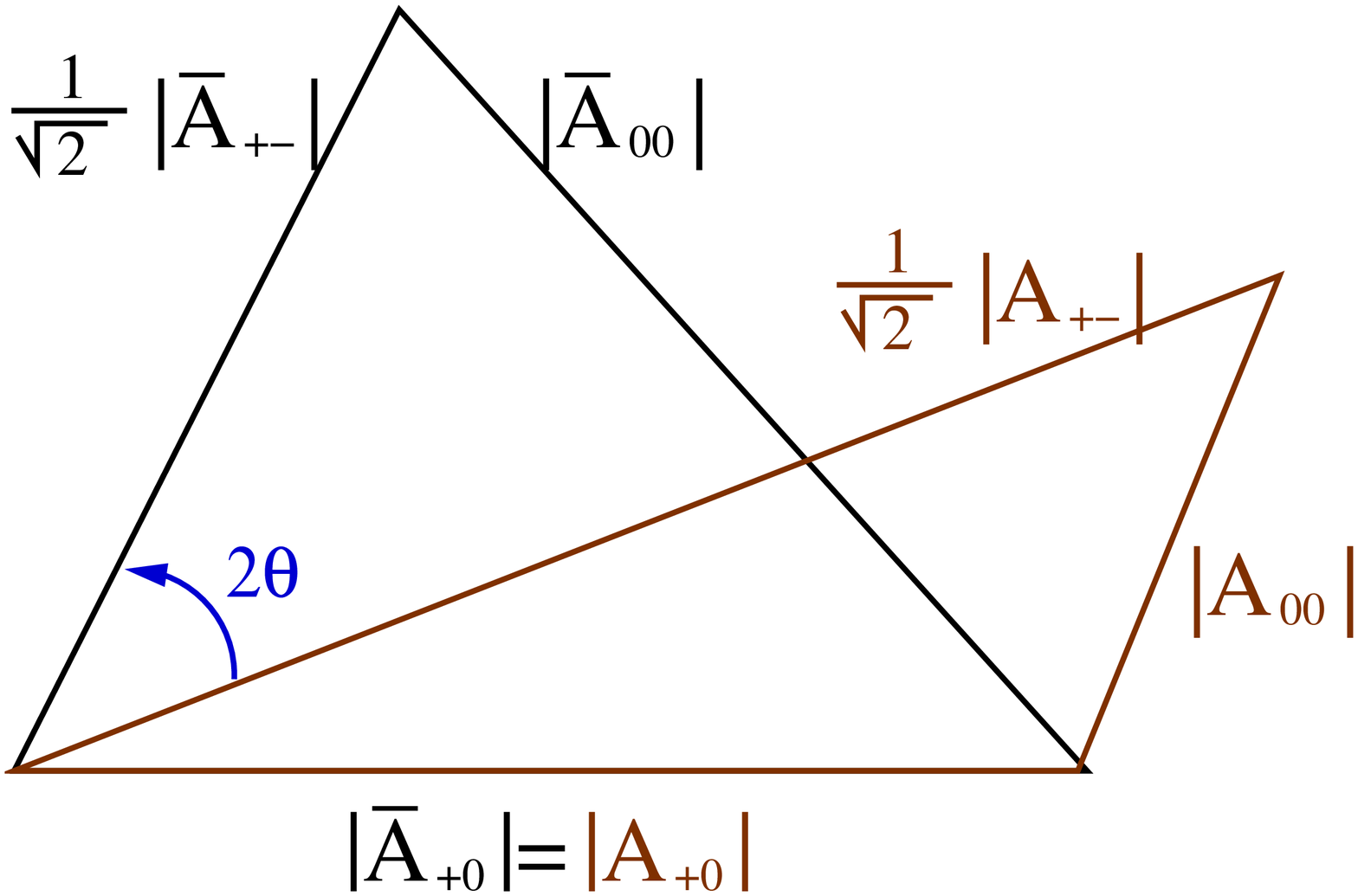}~~~~~~~~~\includegraphics[width=6.1cm]{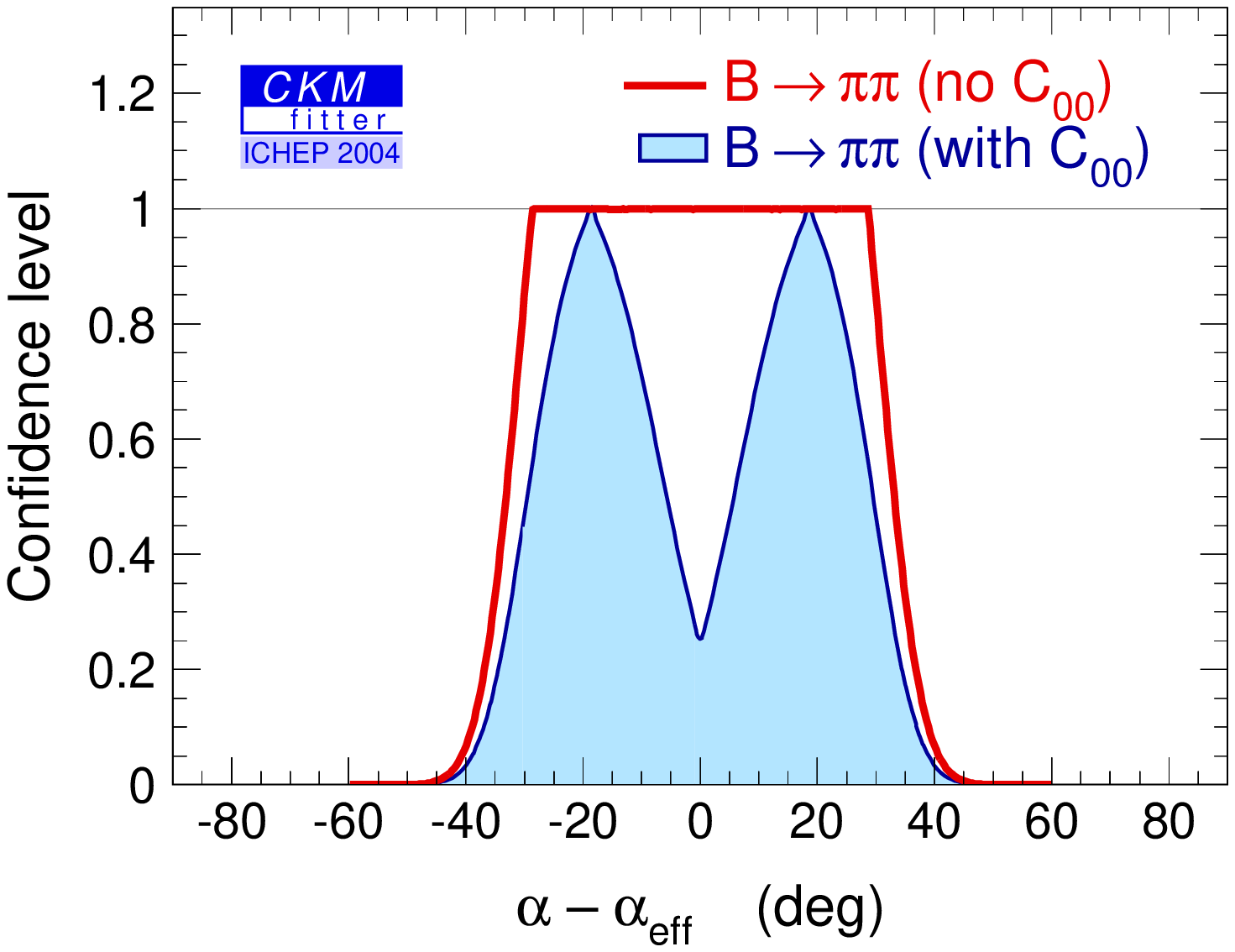}
\end{center}
\caption{
Left: presentation of Eqs. (\ref{Zupantriangle-rel}), (\ref{Zupantree-rel}),
due to Gronau and London \cite{Gronau:1990ka}. 
Only one of four possible triangle orientations is shown. Right: constraints on 
$\alpha -\alpha_{\rm eff}$ from isospin analysis \cite{Ligeti:2004ak}.}  
\end{figure}

The difficulty of this approach is the need to distinguish between $B^0\to \pi^0 \pi^0$ and 
$\bar B^0\to \pi^0\pi^0$ decays, i.e. the need to measure the sides $A_{00}$ and $\bar A_{00}$ of the triangle relations (see 
Fig. 1).
Since the summer of 2004 at least a preliminary isospin analysis is possible, 
as first measurements of $B^0(\bar B^0)\to \pi^0\pi^0$ rates became available,  
$C_{00}=-0.28 \pm 0.39$ and $Br(B\to \pi^0\pi^0)=(1.51\pm 0.28) \cdot 10^{-6}$ \cite{Abe:2004mp}.
Taking a simple weighted average of Belle and BaBar results on $B(t)\to \pi^+\pi^-$ (see 
Table 1),
 the isospin analysis would at present lead to the
constraint on $\alpha -\alpha_{\rm eff}$ shown on Fig. 1
\cite{Ligeti:2004ak}. One does see two 
emerging peaks when information on $A_{00},\bar A_{00}$ is included, however more data is needed 
to constrain $\alpha -\alpha_{\rm eff}$. At present
\begin{equation}
|\alpha -\alpha_{\rm eff}|< 39^\circ  \qquad(90 \% {\rm ~CL}).
\end{equation}
Furthermore the interpretation of $\alpha$ is far from clear due to marginal consistency of $S_{\pi\pi}$ measurements, 
Table 1. 
Recently it was also noted that $\alpha_{\rm eff}> \alpha$, if the magnitude and phase of penguin contributions
is not too large \cite{Gronau:2004sj}.


\begin{table}[t]
\label{Zupantab:pipi}
\begin{center}
\begin{tabular}{|l||c|c|}  
\hline
$B\to\pi^+\pi^-$  &  $\sin2\alpha_{\rm eff}$  &  $C_{\pi^+\pi^-}$ 
\\ \hline\hline
BABAR \cite{Aubert:2004kw} &  $-0.30 \pm 0.17$  &  $-0.09 \pm 0.15$ \\
BELLE \cite{Abe:2004us} &  $-1.00 \pm 0.22$  &  $-0.58 \pm 0.17$ \\
average  &  $-0.61 \pm 0.14$  &  $-0.37 \pm 0.11$ 
\\ \hline
\end{tabular}
\end{center}
\caption{Experimental values of observables in $B\to \pi^+\pi^-$.}
\end{table}

Let us now return to the question of theoretical uncertainties in the isospin analysis \cite{Ciuchini2003}.
There are two sources of isospin breaking: (i) $d$ and $u$ charges are different, and (ii)
$m_u$ does not equal $m_d$. The difference of the light quark charges results in additional operators,
the electroweak penguin operators, in the effective weak hamiltonian. Including EWP in the analysis will not affect the
separate triangle relations (\ref{Zupantriangle-rel}), but only the additional relation (\ref{Zupantree-rel}), 
since $A_{0+}$ no longer receives only tree contributions. Remarkably enough, there exist a relation \cite{Neubert:1998pt,
Gronau:1998fn}
\begin{equation}
H_{\rm eff,~EWP}^{\Delta I =3/2}=-\frac{3}{2}\frac{C_9+C_{10}}{C_1+C_2}
\left|\frac{V_{tb}^*V_{td}}{V_{ub}^*V_{ud}}\right| H_{\rm eff,~c-c}^{\Delta I =3/2}
\end{equation}
which makes the inclusion of EWP fairly straightforward. Instead of Eq. (\ref{Zupantree-rel}) one has
\begin{equation}
e^{i\gamma} A_{+0}=e^{-i(\gamma+{ \delta})} \bar A_{+0},
\end{equation}
where  $\delta \sim 1.5^\circ$ \cite{Gronau:1998fn}. The only assumption that entered this estimate
is the dominance of  EWP operators $Q_{9,10}$, while no estimate of matrix elements is needed. Note that
still $|A_{+0}|= |\bar A_{+0}|$. Deviations from this relations would therefore not test the presence of EWP but 
only the size 
of the Wilson coefficient suppressed EWP operators  $Q_{7,8}$. Note as well, that the same relation 
$e^{i\gamma} T=e^{-i(\gamma+{ \delta})} \bar T$ holds also 
for $\Delta I=3/2$ (tree) amplitudes in the $\rho\rho$ and $\rho\pi$ systems.

Nonzero $m_u- m_d$ difference results in $\pi^0-\eta-\eta'$ mixing, i.e.
$\pi^0$ wave function has small $\eta, \eta'$ admixtures.  Because of this Gronau-London 
triangle relations (\ref{Zupantriangle-rel}) no longer hold \cite{Gardner:1998gz}.
Gardner \cite{Gardner:1998gz} found that this typically leads to $\Delta \alpha \sim 5^\circ$ shift in the extracted
value of $\alpha$. Since we now have more experimental data about the $\pi\pi$ system it would be interesting
to reevaluate this effect, especially if the analysis is extended beyond factorization that was used in 
\cite{Gardner:1998gz}. The analysis of \cite{Gardner:1998gz} also showed that  $\Delta \alpha$
depends on the value of $P/T$ and will thus be different for $\rho\rho$, $\rho\pi$ systems, where 
no such quantitative analysis exists at present.

\subsection{Measurement of $\alpha$ from $B\to \rho \rho$}
The isospin analysis in $B\to \rho\rho$ follows the same lines as for $B\to \pi \pi$, but with
three separate isospin relations (\ref{Zupantriangle-rel}), one for each polarization. However, longitudinally polarized 
final state
 dominates the other two,
$f_L^{+-}=0.99 \pm 0.03 ^{+0.04}_{-0.03}$ \cite{Aubert:2004zr} and 
$f_L^{+0}=0.97^{+0.03}_{-0.07}\pm0.04$ \cite{Aubert:2003mm}. This simplifies the analysis as there is effectively only
one isospin relation. Another difference from the $\pi\pi$ system is that $\rho$ resonances have a nonnegligible decay 
width.
The invariant mass measured from the decay products can thus differ from the pole mass of the $\rho$ resonance.
The two $\rho$ resonances in the final state can therefore also form an $I=1$ state, if
the respective invariant masses are different \cite{Falk:2003uq}. 
This affects the analysis at 
 $O(\Gamma_\rho^2/m_\rho^2)$. As shown in \cite{Falk:2003uq} it is possible to constrain this effect
 experimentally by making 
different fits to the mass distributions.

An ingredient that makes the $\rho\rho$ system favorable over $\pi\pi$ is a small penguin pollution, as can be inferred
from the bound
$Br(B\to \rho^0\rho^0)< 1.1 \cdot 10^{-6} (90 \% {\rm ~CL})$ \cite{Dallapiccola} (cf. also Fig. 1). This gives a measurement
of $\alpha$  from $S_{\rho^+\rho^-}$ using isospin analysis \cite{Dallapiccola}
\begin{equation}
\alpha=[96\pm10\pm4\pm{11}]^\circ,
\end{equation}
with the last error representing the ambiguity due to the presence of
penguins. In obtaining the above result isospin breaking effects, EWP, non-resonant and $I=1$ contributions were neglected.

\subsection{$B\to \rho\pi$}
Since $\rho^\pm \pi^\mp$ are not CP eigenstates, extracting $\alpha$ from this system is more complicated. There are
essentially two approaches, (i) either one uses the full $B(t) \to \pi^+\pi^-\pi^0$ Dalitz plot together with isospin
 \cite{Snyder:1993mx}, or (ii) one uses only the $\rho^\pm\pi^\mp$ region together with 
SU(3) related modes \cite{Gronau:2004tm}.

If the full $B\to \pi^+\pi^-\pi^0$ Dalitz plot is used, one needs to 
model the Dalitz plot, for instance as a fit to a sum of Breit-Wigner forms
\begin{equation}
f(B^0\to 3\pi)=BW(s_+) \underbrace{A(B^0\to \rho^+ \pi^-)}_{A_{+}}+BW(s_-) \underbrace{A(B^0\to \rho^- \pi^+)}_{A_{-}}
+BW(s_0) \underbrace{A(B^0\to \rho^0 \pi^0)}_{A_{0}},
\end{equation}
where for simplicity  only $\rho$ resonances were kept in the sum, but other resonance can be added. From 
time dependent $B(t)\to 3 \pi$ Dalitz plot analysis one has 27 real observables, of which 18 measure the interference
between different $\rho$ resonance bands.
In this way it is
possible to determine $A_{\pm,\;0}$, $\bar A_{\pm,\;0}$, up to an overall phase,
i.e. there are 11 independent measurables. A potential problem can arise from the fact 
that the peaks of $\rho$ resonance bands do not overlap,
but are separated by approximately one decay width. To measure the 11 observables correctly one therefore has to
model the tails of the resonances correctly.

In order to extract $\alpha$ from $A_{\pm,\;0}$, $\bar A_{\pm,\;0}$ additional input is needed.
First let us define tree and penguin contributions according to whether or not they contain CKM weak phase
\begin{equation}
\label{Zupantree-peng-decomp}
A_\pm=e^{i \gamma} t_{\pm}+p_{\pm},\qquad\qquad A_0=e^{i \gamma} t_{0}+p_{0},
\end{equation} 
and similarly for $\bar{A}_{\pm},\bar{A}_{0}$, but with a sign of $\gamma$ flipped. The $\Delta I=3/2$ part of 
the weak hamiltonian has 
a CKM phase, so the penguins $p_{\pm,0}$ are purely $\Delta I=1/2$ (neglecting EWP). This leads to 
an isospin relation \cite{Snyder:1993mx,Lipkin:1991st}
\begin{equation}
\label{p-relation}
p_{0}=-\tfrac{1}{2}(p_++p_-),
\end{equation}
which reduces the number of unknowns to 10. One possible choice of unknowns is
$\alpha$, $|t_\pm|$, $|t_0|$ $\arg t_\pm$, $|p_\pm|$, {$\arg p_\pm$}. There is thus 
 enough information to determine all of them. Explicitly, the observable that gives $\alpha$ directly is
\begin{equation}
A_{+-}+A_{-+}+2 A_{00}=T\Rightarrow-\Im\left(\tfrac{q}{p}\tfrac{\bar T}{T}\right)=\sin(2\alpha).
\end{equation}
There are some further comments that apply to the Snyder-Quinn method. As already stated, the 
effects due to isospin breaking have not been analysed quantitatively yet. However,
isospin breaking  will enter only in relation (\ref{p-relation}). Since penguins are small,
 $|p_\pm/t_\pm|\sim 20\%$ \cite{Gronau:2004tm},
it is reasonable to expect that isospin breaking effects will also be small, or at least smaller
than in the $B\to \pi\pi$ case. In addition,  if $A_{5/2}\ne 0$, only the part of $A_{5/2}$ that
 has the same weak phase as $p_{\pm,0}$ will affect the analysis by modifying the relation in Eq. (\ref{p-relation}).
These contributions would come from electromagnetic final state rescatering
of penguin contributions, leading to negligible effect. 

BaBar performed the Snyder-Quinn analysis (but keeping 10 out of 27 observables fixed to zero), obtaining 
\cite{Aubert:2004iu}
\begin{equation}
\alpha=(113^{+27}_{-17}\pm6)^\circ.
\end{equation}
Note that there is only one solution in $[0^\circ,180^\circ)$. 

The potential problem of having to model the tails of the $\rho$ bands can be avoided by using 
just the $\rho^\pm \pi^\mp$ final state and the $SU(3)$ related modes \cite{Gronau:2004tm}. As in 
 (\ref{Zupantree-peng-decomp}),
the tree and penguin 
contributions are defined according to their weak phases. In total there are 
8 unknowns: $|t_\pm|$, $|p_\pm|$, $\arg\big(p_\pm/t_\pm\big)$, $\arg\big(t_-/t_+\big)$, $\alpha$, but just 6 observables.
Additional infromation on penguin contributions can be obtained from SU(3) related
$\Delta S=1$ modes, in which penguins are CKM enhanced
and tree terms CKM suppressed compared to the $\rho^\pm \pi^\mp$ final state. Since penguin contributions are
small, the error introduced
 because of the SU(3) breaking will not be large. Note that in order to relate the $\Delta S=1$ and $\Delta S=0$ channels, 
annihilation like topologies were neglected.

To resolve ambiguties an additional assumption of  $\arg(t_-/t_+)$ being smaller than $90^\circ$
had to be used. This leads to 
\begin{equation}
\alpha=(94 \pm 4\pm {15})^\circ
\end{equation}
with the last error the combined error coming from $\alpha-\alpha_{\rm eff}$ difference and the estimate of
SU(3) breaking effects. To obtain this number
no interference information was used (i.e. experimental data from both BaBar \cite{Aubert:2004iu} and Belle 
\cite{Wang:2004va}
was used).  Also, only 
bounds on penguins were used, not a complete SU(3) fit. 
In the future an unconstrained fit to obtain $\alpha$ could be performed. This would lead to a single 
solution for $\alpha$, with all ambiguities resolved. As already stated, the SU(3) breaking on extracted $\alpha$ 
would be  small, of order
$p_\pm^2/t_\pm^2$. A Monte Carlo study with up to $30\%$ SU(3) breaking on
penguins for instance gives $\sqrt{\langle(\alpha^{\rm out}-\alpha^{\rm in})^2\rangle}\sim 2^\circ$ \cite{Gronau:2004tm}.

\section{Measuring $\gamma$}
\subsection{$B^\pm\to D K^\pm$}
There are many methods that fall into this class, all of which use the interference between 
$b\to c \bar{u} s$ and $b\to u \bar{c} s$ \cite{Gronau:1991dp}.
In the case of charged $B$ decays this means that the interference is between 
$B^-\to D K^-$ followed by $D\to f$ decay and 
$B^-\to \bar{D} K^-$ followed by $\bar{D}\to f$, where 
 $f$ is any common final state of $D$ and $\bar{D}$. What makes this method very powerful is that 
there are no penguin contributions and therefore almost no theoretical uncertainties, with all the
hadronic unknowns in principle obtainable from experiment (with problems in measuring color 
suppressed $B^-\to \overline{D^0} K^-$ decay \cite{Atwood:1996ci}).

Different methods can be grouped according to the choice of the final state $f$, which can be (i)
a CP- eigenstate (e.g. $K_S \pi^0$) \cite{Gronau:1991dp}, (ii)
a flavor state (e.g. $K^+\pi^-$) \cite{Atwood:1996ci} , (iii)
a singly Cabibbo suppressed (e.g. $K^{*+} K^-$) \cite{Grossman:2002aq} or (iv) 
a many-body final state (e.g. $K_S\pi^+\pi^-$) \cite{Giri:2003ty}. There are also other extensions:
many body $B$ final states (e.g. $B^+\to D K^+\pi^0$) \cite{Aleksan:2002mh}, 
 $D^{0*}$ in addition to $D^0$, self tagging $D^{0**}$ \cite{Sinha:2004ct} or 
neutral $B$ decays (time dependent and time-integrated) can be used \cite{Gronau:2004gt,Kayser:1999bu}.

In this talk we focus on extracting $\gamma$ from $B^\pm\to (K_S\pi^+\pi^-)_D K^\pm$, since this is 
experimentally most advanced. Both experiments use $D^*$ and $D$ decays, where a 
subtelty of a sign flip in the use of $D^*$
 has been pointed out only recently \cite{Bondar:2004bi}. The BaBar result \cite{Aubert:2004kv}
\begin{equation}
\gamma=(88\pm41\pm19\pm10)^\circ,
\end{equation}
should therefore be treated as preliminary only. Belle on the other hand obtains \cite{Gershon_talk}
\begin{equation}
\gamma=(68^{+14}_{-15}\pm13\pm11)^\circ.
\end{equation}
Note that only a single solution for $\gamma$ is obtained in $[0, 180)^\circ$ range.

For details on how the method works see \cite{Giri:2003ty,Aubert:2004kv,Gershon_talk,Poluektov:2004mf}. 
We will just make several statements regarding the remaining theoretical
errors. First of all, it is possible to extend this approach beyond Breit-Wigner fits of Dalitz plot, so that
there is no modeling error left \cite{Giri:2003ty,Atwood:2003mj}. Also, the 
effect of $D-\bar D$ mixing is included automaticaly, if $D^*$ tagged $D$ decays are used to measure the observables of 
$D$ system.\footnote{I thank T. Gershon for pointing this out.} The largest remaining theoretical error is due to possible
direct CP violation in the $D$ decay, which is, however, highly CKM suppressed by $\lambda^5\sim 5\cdot 10^{-4}$. 
The measurement of $\gamma$ will therefore be dominated by experimental errors for years to come.


\subsection{$\gamma$ from $B(t)\to D^{(*)+}D^{(*)-}$}
This is a very recent method \cite{Datta:2003va}. Again, the amplitude is 
split into tree and penguin according to CKM
\begin{equation}
A_D=A(B^0\to D^+D^-)=\underbrace{t}_{V_{cb}^*V_{cd}}+\underbrace{p e^{i\gamma}}_{V_{ub}^* V_{ud}}
\end{equation}
Value of $t$ is determined from $B^0\to D_s^+D^-$ using SU(3) with leading SU(3) breaking correction accounted for
$
\frac{t'}{t}=\frac{V_{cs}}{V_{cd}}\frac{f_{D_s}}{f_D}$, with  
 subleading corrections estimated to be below $5\%-10\%$. 
At present additional approximations are needed to obtain bounds on $\gamma$ from $D^{*+}D^-$, $D^{*+}D^{*-}$. This 
leads to three viable regions for $\gamma$ at $68\%$ CL, $\gamma\in[19.4^\circ, 80.6^\circ] $ or
$\gamma\in[120^\circ, 147^\circ] $
or 
$\gamma\in[160^\circ, 174^\circ]$.

\subsection{$\sin(2\beta + \gamma)$}
The combination $\sin(2\beta +\gamma)$ can (at least in principle) be extracted 
very cleanly from the time dependent measurement
$B(t)\to D^{(*)}\pi^-$ \cite{Dunietz:1997in}. Until the small direct CP asymmetry is measured,
however, the
weak phase $\gamma$ and the strong phase $\delta$ can be extracted from $
S_{D^{(*)\pm} \pi^\mp}= {2 r}/({1+r^2})\cdot \sin(2 \beta +\gamma\pm\delta)$
only, if the ratio of the two interfering amplitudes
$r=|A(B^0\to D^{(*)+}\pi^-)/A(\bar B^0\to D^{(*)+}\pi^-)| $ is known. 
This ratio can be obtained using $SU(3)$ from $Br(B^0\to D_s^{(*)+}\pi^-)$.
Assuming factorization, taking $f_{D_s^{(*)}}/f_{D^{(*)}}$ from lattice, and neglecting
(very) small anihhilation like diagrams, this gives  
$
r_{D\pi}=0.019\pm0.04 $, $r_{D^*\pi}=0.017^{+0.005}_{-0.007}$.
Using this number BaBar obtains 
$|\sin(2 \beta +\gamma)|>0.58 ~(90\% {\rm ~CL})$ from partially reconstructed $B\to D^{*\mp}\pi^\pm$ \cite{Aubert:2004pt}.

\section{Conclusions}
In conclusion, we have working tools to determine angles $\alpha$ and $\gamma$  of the CKM unitarity triangle. 
The experimental
situation looks much more favorable than expected a few years ago. For instance, measurements of $\alpha$ 
are already reaching precision level, where one has to start worrying about theoretical errors.\\[5mm]
{\Large {\bf Acknowledgements}}\\[2mm]
I thank Y. Grossman and M. Gronau for carefully reading the manuscript.

%
\label{ZupanEnd}

\end{document}

%% file: fpcp04-macro.tex
\def\beq{\begin{equation}}
\def\eeq#1{\label{#1}\end{equation}}
\def\eeqn{\end{equation}}

\def\beqa{\begin{eqnarray}}
\def\eeqa#1{\label{#1}\end{eqnarray}}
\def\eeqan{\end{eqnarray}}

\let\bar=\overbar

\def\Dslash{\not{\hbox{\kern-4pt $D$}}}
\def\dslash{\not{\hbox{\kern-2pt $\del$}}}

\def\msb{{\bar{\ssstyle M \kern -1pt S}}}

\def\BB0bar{B^0 {\overline B}^0}
\def\BB0dbar{B_d^0 {\overline B}_d^0}
\def\BB0sbar{B_s^0 {\overline B}_s^0}

\RequirePackage{xspace}

\usepackage{relsize}
\def\babar{\mbox{\slshape B\kern-0.1em{\smaller A}\kern-0.1em
    B\kern-0.1em{\smaller A\kern-0.2em R}}}

\def\Kbar  {\kern 0.2em\overline{\kern -0.2em K}{}\xspace}

\def\Kz    {\ensuremath{K^0}\xspace}
\def\Kzb   {\ensuremath{\Kbar^0}\xspace}
\def\KzKzb {\ensuremath{\Kz \kern -0.16em \Kzb}\xspace}
\def\Kp    {\ensuremath{K^+}\xspace}
\def\Km    {\ensuremath{K^-}\xspace}

\def\KpKm  {\ensuremath{\Kp \kern -0.16em \Km}\xspace}

\def\Dbar    {\kern 0.2em\overline{\kern -0.2em D}{}\xspace}

\def\Dz      {\ensuremath{D^0}\xspace}
\def\Dzb     {\ensuremath{\Dbar^0}\xspace}
\def\DzDzb   {\ensuremath{\Dz {\kern -0.16em \Dzb}}\xspace}
\def\Dp      {\ensuremath{D^+}\xspace}
\def\Dm      {\ensuremath{D^-}\xspace}

\def\DpDm    {\ensuremath{\Dp {\kern -0.16em \Dm}}\xspace}

\def\Bbar    {\kern 0.18em\overline{\kern -0.18em B}{}\xspace}

\def\BB      {\ensuremath{B\Bbar}\xspace} 
\def\Bz      {\ensuremath{B^0}\xspace}
\def\Bzb     {\ensuremath{\Bbar^0}\xspace}
\def\BzBzb   {\ensuremath{\Bz {\kern -0.16em \Bzb}}\xspace}
\def\Bu      {\ensuremath{B^+}\xspace}
\def\Bub     {\ensuremath{B^-}\xspace}

\def\BpBm    {\ensuremath{\Bu {\kern -0.16em \Bub}}\xspace}

\mathchardef\Upsilon="7107
\def\Y#1S{\ensuremath{\Upsilon{(#1S)}}\xspace}

\mathchardef\Deltares="7101
\mathchardef\Xi="7104
\mathchardef\Lambda="7103
\mathchardef\Sigma="7106
\mathchardef\Omega="710A

\def\Deltabar{\kern 0.25em\overline{\kern -0.25em \Deltares}{}\xspace}
\def\Lbar{\kern 0.2em\overline{\kern -0.2em\Lambda\kern 0.05em}\kern-0.05em{}\xspace}
\def\Sigbar{\kern 0.2em\overline{\kern -0.2em \Sigma}{}\xspace}
\def\Xibar{\kern 0.2em\overline{\kern -0.2em \Xi}{}\xspace}
\def\Obar{\kern 0.2em\overline{\kern -0.2em \Omega}{}\xspace}
\def\Nbar{\kern 0.2em\overline{\kern -0.2em N}{}\xspace}
\def\Xb{\kern 0.2em\overline{\kern -0.2em X}{}\xspace}

\newcommand{\tev}{\ensuremath{\mathrm{\,Te\kern -0.1em V}}\xspace}
\newcommand{\gev}{\ensuremath{\mathrm{\,Ge\kern -0.1em V}}\xspace}
\newcommand{\mev}{\ensuremath{\mathrm{\,Me\kern -0.1em V}}\xspace}
\newcommand{\kev}{\ensuremath{\mathrm{\,ke\kern -0.1em V}}\xspace}
\newcommand{\ev}{\ensuremath{\mathrm{\,e\kern -0.1em V}}\xspace}
\newcommand{\gevc}{\ensuremath{{\mathrm{\,Ge\kern -0.1em V\!/}c}}\xspace}
\newcommand{\mevc}{\ensuremath{{\mathrm{\,Me\kern -0.1em V\!/}c}}\xspace}
\newcommand{\gevcc}{\ensuremath{{\mathrm{\,Ge\kern -0.1em V\!/}c^2}}\xspace}
\newcommand{\mevcc}{\ensuremath{{\mathrm{\,Me\kern -0.1em V\!/}c^2}}\xspace}

\def\mus  {\ensuremath{\rm \,\mus}\xspace}

\def\mus        {\ensuremath{\,\mu{\rm s}}\xspace}    

\def\to                 {\ensuremath{\rightarrow}\xspace}

\def\pep2{PEP-II}

\def\gsim{{~\raise.15em\hbox{$>$}\kern-.85em
          \lower.35em\hbox{$\sim$}~}\xspace}
\def\lsim{{~\raise.15em\hbox{$<$}\kern-.85em
          \lower.35em\hbox{$\sim$}~}\xspace}

\xspace

\def\jetset74   {\mbox{\tt Jetset \hspace{-0.5em}7.\hspace{-0.2em}4}\xspace}
